# SD-REE: A Cryptographic Method to Exclude Repetition from a Message

Somdip Dey
Department of Computer Science
St. Xavier's College [Autonomous]
Kolkata, India.
Email: somdip007@hotmail.com

*Abstract* – **In this paper, the author presents a new cryptographic technique, SD-REE, to exclude the repetitive terms in a message, when it is to be encrypted, so that it becomes almost impossible for a person to retrieve or predict the original message from the encrypted message. In modern world, cryptography hackers try to break a code or cryptographic algorithm [1,2] or retrieve the key, used for encryption, by inserting repetitive bytes / characters in the message and encrypt the message or by analyzing repetitions in the encrypted message, to find out the encryption algorithm or retrieve the key used for the encryption. But in SD-REE method the repetitive bytes / characters are removed and there is no trace of any repetition in the message, which was encrypted.**

## I. Introduction

In modern world, security is a big issue and securing important data is very essential, so that the data can not be intercepted or misused for illegal purposes. For example, we can assume the situation where a bank manager is instructing his subordinates to credit an account, but in the mean while a hacker interpret the message and he uses the information to debit the account instead of crediting it. Or we can assume the situation where a military commander is instructing his fellow comrades about an attack and the strategies used for the attack, but while the instructions are sent to the destination, the instructions get intercepted by enemy soldiers and they use the information for a counter-attack. This can be highly fatal and can cause too much destruction. So, different cryptographic methods [1,2] are used by different organizations and government institutions to protect their data online. But, cryptography hackers are always trying to break the cryptographic methods or retrieve the key by different means and one of such methods include the process of inclusion of repetitive texts or characters in a message and then encrypt them to study the cryptanalysis of the method and break the algorithm or fetch the key. Another common method of cryptanalysis is by observing and analyzing the frequency of characters of encrypted message and retrieve the key by finding a regular pattern. But SD-REE is a cryptographic method, which can be used to exclude the repetitive characters in a message to be encrypted and this technique is a type of symmetric key cryptography.

The modern day cryptographic methods are of two types: (i) symmetric key cryptography, where the same key is used for encryption and for decryption purpose. (ii) Public key cryptography, where we use one key for encryption and one key for decryption purpose.

Symmetric key algorithms are well accepted in the modern communication network. The main advantage of symmetric key cryptography is that the key management is very simple. Only one key is used for both encryption as well as for decryption purpose. There are many methods of implementing symmetric key. In case of symmetric key method, the key should never be revealed / disclosed to the outside world or to other user and should be kept secure.

As discussed earlier, SD-REE is also a symmetric key cryptographic method and the best part of this method is that this method can be used with other cryptographic method to make the encryption strong enough to be hacked by hackers and, while SD-REE method is being used.

In this paper the authors present the cryptographic technique SD-REE, which is a modified form of Advanced Caesar Cipher Cryptographic Method [5,6]. In cryptography, a Caesar cipher, also known as a Caesar's cipher or the shift cipher or Caesar's code or Caesar shift, is one of the simplest and basic known encryption techniques. It is a type of replace cipher in which each letter in the plaintext is replaced by a letter with a fixed position separated by a numerical value used as a "key".

Caesar Cipher is or was probably the very first encryption methodology. It is a type of substitution cipher in which each letter in the plaintext is replaced by a letter some fixed number of positions down the alphabet. For example, with a shift of 3, A would be replaced by D, B would become E, and so on.

In SD-REE, we extract the ASCII value of each byte (character) of the text or file to be encrypted and then we add two numbers, which we generate randomly from the pass-key (symmetric key) provided for encryption, to generate a new ASCII value and print out the character of that ASCII value. The two numbers, which are generated from the pass-key, are generated totally randomized and it is impossible to predict the pass-key for that reason. Again, one of the two numbers is generated using a random polynomial function and it changes every time with the byte, of which ASCII value it is being added to. For this reason, it is almost impossible to predict the pattern of the cipher method.

In SD-REE we take the symmetric key as password given by the user and from that key we generate unique codes, which are successively used to encrypt the message. The methods used in SD-REE to encrypt the message can be reverse engineered to decrypt the same message by providing the key (password).

## II. Encryption

### a) Generation of Code and power_ex from the Symmetric Key

The key is provided by the user in a string format and let the string be ‘pwd[]’. From the given key we generate two numbers:
‘code’ and ‘power_ex’, which will be used for encrypting the message. First we generate the ‘code’ from the pass key.
Generation of code is as follows:

To generate the code, the ASCII value of each character of the key is multiplied with the string-length of the key and with $2^i$, where ‘i’ is the position of the character in the key, starting from position ‘0’ as the starting position. Then we sum up the resultant values of each character, which we got from multiplying, and then each digit of the resultant sum are added to form the ‘pseudo_code’. Then we generate the code from the pseudo_code by doing modular operation of pseudo_code by 16, i.e.

code = (pseudo_code Modulus 16).

If code==0, then we set code =pseudo_code

The Algorithm for this is as follows:

Let us assume, pwd[] = key inserted by user

pp= $2^i$ , i=0,1,2,……..n; $n \in N$.

Note: i can be treated as the position of each character of the key.

Step-1 : p[] = pwd[]
Step-2 : pp = $2^i$
Step-3 : i=0
Step-4 : p[i] = pwd[i];
Step-5 : p[i] = p[i] * strlen(pwd) * pp;
Step-6: csum = csum + p[i];
Step-7: i=i+1
Step-8: if i < length(pwd) then go to step-4
Step-9: if csum $\neq$ 0 then go to Step-10 otherwise go to Step-14
Step-10: c = Mod(csum , 10)
Step-11: pseudo_code=pseudo_code +c;
Step-12: csum = Int(csum / 10)
Step-13: Go to step-9
Step-14: code =Mod (pseudo_code, 16)
Step-15: End

Note: length(pwd)= number of characters of the secret key pwd[].

The ‘power_ex’ is calculated as follows:
We generate power_ex from the pseudo_code generated from the above method. We add all the digits of the pseudo_code and assign it as temporary_power_ex. Then we do modular operation on temporary_power_ex with code and save the resultant as power_ex.
i.e.

power_ex = Mod (temporary_power_ex, code)

If power_ex = 0 OR power_ex = 1, then we set power_ex = code.

For example, if we choose the password, i.e. the key to be ‘hello world’. Then,
Length of pwd = 11
code = 10
power_ex = 4

Thus, we generate code and power_ex from the key provided by the user.

### b) Encrypting the Message using code and power_ex (Modified Caesar Cipher Method)

Now we use the code and power_ex, generated from the key, to encrypt the main text (message). We extract the ASCII value of each character of the text (message to be encrypted) and add the code with the ASCII value of each character. Then with the resultant value of each character we add the $(\text{power_ex})^i$, where i is the position of each character in the string, starting from ‘0’ as the starting position and goes up to n, where n=position of end character of the message to be encrypted, and if position = 0, then $(power_ex)^i = 0$. It can be given by the formula:

$$\textbf{text[i] = text[i] + code + (power_ex)}^{i}$$

If, ASCII value of text[i] > 255, then set
text[i] = (text[i] Modulus 256)
Note: 'i' is the position of each character in the text and text[] is the message to be encrypted, where text[i] denotes each character of the text[] at position 'i'.

For example, if the text to be encrypted is 'aaaa' and key=hello world, i.e. text[]=aaaa and pwd=hello world, then,
$a^0$ -> 97 + 10 + 0 = 107 ->k
$a^1$ -> 97 + 10 + 4 = 111 ->o
$a^2$ -> 97 + 10 + 16 = 123 ->{
$a^3$ -> 97 + 10 + 64 = 171 -> «
where 0-3 are the positions of 'a' in text[]
(as per formula given above)
The text 'aaaa' becomes 'ko{«' after execution of SD-REE method.

Since, the value of $(power_ex)^i$ increases with the increasing number of character (byte) i.e. with the increasing number of string length, so we have applied the method of **Modular Reduction** [3,4] to reduce the large integral value to a smaller integral value.

To apply Modular Reduction we apply the following algorithm:
Step 1: n = power_ex * code * 10 ; generate a random number 'n' from code and power_ex
Step 2: calculate $n^{th}$ prime number
Step 3: i=0
Step 4: $(power_ex)^i = Mod((power_ex)^i, (n^{th}$ prime number))
Step 5: i=i+1
Step 6: if i<length(text) then go to step-4
Step-7: End

Following the above step, we can reduce the value of $(power_ex)^i$ to a significantly smaller usable number.

## III. Decryption

The above processes of encryption of SD-REE method can be reverse engineered to get back the original text (message) and thus decryption of the encrypted text can be executed.

The decryption method mainly works on the formula:

$$text[i] = text[i] - code - (power_ex)^{i}$$

Note: If (ASCII Value of text[i] < 0) condition holds then set:
text[i] = Mod( text[i], 256 )

## IV. Results and Discussions

In the following table few results are given:

<table>
<tr><th>Text to be Encrypted</th><th>Encrypted Text</th></tr>
<tr><td>aaaaaaaa</td><td>qtzŒÂeM%5</td></tr>
<tr><td>111aaaaaaaa111</td><td>AHrÉÛ<br>`ù~o®?„</td></tr>
<tr><td>bbbcccddd</td><td>dht¥fiv¦$4</td></tr>
<tr><td>1111155555</td><td>59Eu6=Iy:%0</td></tr>
<tr><td>aaa2277bbccdd</td><td>egi>F[{æ gikpx</td></tr>
<tr><td>Hello AAABBBCC CDDDEEER get well at 11111111 22222222</td><td>QwÆQ3¹_õTœ'Üa øU®Ðç[)pn})□nuu jzôÊO÷åC õËPøæC›</td></tr>
<tr><td>ÿÿÿÿÿÿÿÿÿÿ</td><td>ÓÖ'=`ÁWßÌÌ©¨Á¢O</td></tr>
<tr><td>旑旑旑旑</td><td>ï©ëÊZ*<br>]E÷ðu</td></tr>
<tr><td>Somdip BBBBBBBBnnnnb bbbb</td><td>\□ÇI-<br>%A>%9öTœ'%6Û`5#□<br>×í%5k</td></tr>
<tr><td>The Xaverian vision of education encompasses a sensitive understanding of the realities defining our existence. It is our committed mission to incorporate in the minds of our students an awareness of this social vision so that they can be better equipped to take up the larger responsibilities in every sphere of life.</td><td>^z¯,rë„ }<br>{«z:w□t□¸,‰ð*o<br>n□mk~syx*<br>oxmywzk}}o}*k*}o<br>x}s~s□o*□xno|}~k<br>xnsxq*yp*<br>~ro*|okvs~<br>so}*nopsx<br>sxq*y□|*o,s}~oxmo8<br>*S~*s}*y□|*mywws~<br>~on*ws<br>}}syx*~y*sxmy|zy|k~o<br>*sx*~r<br>o*wsxn}*yp*y□|*}~<br>□nox~}*<br>kx*k□k|oxo}}*yp*~<br>rs}*}ym<br>skv*□s}syx*}y*~rk~*~<br>r<br>oƒ*m<br>kx*lo*lo~~o|*o<br>{□szzon*~y*~<br>kuo*□z*~ro*vk|qo|<br>*|o}zyx}sl<br>svs~so}*sx*o□o|ƒ*}zro|<br>o<br>*yp*vspo8</td></tr>
</table>

From the above results, it can be seen that, all the repetitive terms are excluded from the encrypted text and it can never be figured out just from the encrypted text that there was any repetition in the text message. SD-REE is a provable good method to exclude repetitive terms, but, it is also evident from the results shown above, that SD-REE is not a very good cryptographic method to encrypt a full text, i.e. it is not a full proof encryption technique. And this problem can be avoided by adding other encryption techniques with this method.

**Discussions and Limitations of SD-REE:**

SD-REE method should never be treated as a lone method for encryption, but this method is a cryptographic method, which exclude the repetitive characters from the text message, which is to be encrypted. SD-REE is not meant to be used alone as a whole encryption technique, but this method should be used in other cryptographic method to make those method strong.

In most of the recent encryption techniques, symmetric key, which is used for encryption, is mostly 16 –bits long. So, for SD-REE method, if a 16-bit key is provided for encryption and if the ASCII value of the bit is 255,then,

$$csum = \sum_{i=1}^{16} 255 * i * (2)^i$$
$$= 2404388880,$$

then, code = 45 and
power_ex = 9 [According to SD-REE method]

And if the number of bits in the symmetric key increases, then there is a possibility that the value of power_ex may increase too, and this is a problem. If the value of power_ex is very big then the value of $(power_ex)^i$ will be even larger, and since this value is used in SD-REE encryption method, so the value may be very large to store in memory. This problem may arise if SD-REE method is used to encrypt large text files or any other type of large files.

So, to do away with this problem, we have included the step in ‘power_ex’ generation:
power_ex = (temporary_power_ex Modulus code);
If power_ex == 0 OR power_ex == 1, then we set power_ex = code.
Note: To do away with the above stated problem the author have provide this provision; although, this problem can be handled by using modular reduction to reduce the large integral value to suitable form to compute the SD-REE method.

But, this provision may be changed as per user or the cryptographer, and is always welcomed to change this part as needed.

## V. Spectral Analysis And Cryptanalysis

One of the classical cryptanalysis method used, is by detecting the frequency of characters in the encrypted text (message). So to test the effectiveness of this method, spectral analysis of the frequency of characters are closely observed.

Using this method we ran many analysis and tested different strings as input and used various methods of cryptanalysis. To show the usefulness and integrity of this cryptographic module, we used spectral analysis of the frequency of characters.

We chose few special test cases and observed the spectral analysis of the frequency of characters of the encrypted text using SD-REE:

<table>
<tr><th>Message</th><th>Encrypted Messag</th></tr>
<tr><td>aaaaaaaaaaaaaaaaaaaaa<br>aaaaaaaaaaaaaaaaaabbb<br>bbbbbaaaaaaaaaaaaaaa<br>aaaaaaaaaaaaaaaaaaaaa<br>aa (Palindrome of 76 ‘a’ and 8 ‘b’)</td><td>@X²ÈÒÄnm¥□¢``’<br>68F@Ùq□=qxqÈqäqìq<br>Tq`q<br>q8qArÚr□ùür<br>———————————<br>-<br>"%(+.147:=@CFILOR<br>UX[^adgjmq</td></tr>
<tr><td>(128 times ASCII Value (1) )</td><td>□□<br><br>□#C□U§□□a□-<br>;s5g□7k%G□e□/[□<br><br>□#C□U§□□a□-<br>;s5g□7k%G□e□/[□<br><br>□#C□□□□<br>□#□□□§□□□□□;□<br>5□□□k□G□e□/□□□<br>□#□□□§□□□□□;□<br>5□□□k□G□e□/□□□<br>□#□□</td></tr>
<tr><td>aaaaaaaaaaaaaaaaaaaaa<br>aaaaaaaaaaaaaaaaaaaaa<br>aaaaaaaaaaaaaaaaaaaaa<br>aaaaaaaaaaaaaaaaaaaaa<br>aaaaaaaaaaaaaaaaaaaaa<br>aaaaaaaaaaaaaaaaaaaaa<br>aaaaaaaaaaaaaaaaaaaaa<br>aaaaaaaaaaaaaaaaaaaaa<br>aaaaaaaaaaaaaaaaaaaaa<br>aaaaaaaaaaaaaaaaaaaaa<br>aaaaaaaaaaaaaaaaaaaaa<br>aaaaaaaaaaaaaaaaaaaaa<br>aaaaaaaaaaaaaaaaaaaaa<br>aaaaaaaaaaaaaaaaaaaaa<br>aaaaaaaaaaaaaaaaaaaaa<br>aaaaaaaaaaaaaaaaaaaaa<br>aaaaaaaaaaaaaaaaaaaaa<br>aaaaaaaaaaaaaaaaaaaaa<br>aaaaaaaaaaaaaaaaaaaaa<br>aaaaaaaaaaaaaaaaaaaaa</td><td>qx¢ÈÒä^ìfT¢`’<br>68F@Ùq□=qxqÈqäqìq<br>Tq`q<br>q8q@qÙq□øûþ<br>———————————<br>-<br>"%(+.147:=@CFILOR<br>UX[^adgjmpsvy|□,…<br>ˆ‹□‘”—<br>š□ £¦©¬¯²µ¸»¾ÁÄÇÊÍ<br>ÐÓÖÙÜßâåèëîñô÷úý4<br>———————————<br>-!$'*-<br>0369<?BEHKNQTWZ<br>]`cfilorux{~□„‡Š<br>□□“–<br>™œŸ¢¥¨«®±´·º½ÀÃ<br>ÆÉÌÏÒÕØÛÞáäçêíðóö</td></tr>
</table>

<table>
<tr>
<td>aaaaaaaaaaaaaaaaaaaaa<br>aaaaaaaaaaaaaaaaaaaaa<br>aaaaaaaaaaaaaaaaaaaaa<br>aaaaaaaaaaaaaaaaaaaaa<br>aaaaaaaaa (512 times<br>‘a’)</td>
<td>ùüÿ<br>#&),/258;>ADGJMPS<br>VY_behknqtwz}□ƒ†<br>‰Œ□’•˜›□¡¤§ª-º³¶¹¼¿<br>ÂÅÈËÎÑÔ×ÚÝàãæéìïò<br>õøûþ<br>________________<br><br>-<br>"%(+.147:=@CFILOR<br>UX[^adgjmpsvy|□,…ˆ<br>‹□‘”—<br>š□ £¦©¬¯²µ¸»¾ÁÄÇÊÍ<br>ÐÓÖÙÜßâåèëîñô÷úý5<br>________________<br>-!$'*-<br>0369<?BEHKNQTWZ<br>]`cfilorux{~□„‡Š□□“<br>–<br>™œŸ¢¥¨«®±´·º½ÀÃ<br>ÆÉÌÏÒÕØÛÞáäçêíðóö<br>ùüÿ<br>#&),/258;>ADGJMPS<br>VY_behknq</td>
</tr>
</table>

Test case 1 shows the encryption of a palindrome, which consists of 76 ‘a’ and 8 ‘b’, using SD-REE. Total time taken for encryption is 1 second. Fig 1.1 shows the spectral analysis of the frequency of characters of the palindrome, and Fig 1.2 shows the spectral analysis of the frequency of characters after the message being encrypted and repetition being excluded after the execution of SD-REE method.

Test case 2 shows the encryption of ASCII Value (1) 128 times, which is itself another palindrome. Total time taken for encryption is 2 seconds. Fig 2.1 shows the spectral analysis of the frequency of characters of the text message and Fig 2.2 shows the spectral analysis of the frequency of characters of the encrypted text.

Test case 3 shows the encryption of a text which contains ‘a’ 512 times. Total time taken for encryption is 2 seconds. Fig 3.1 shows the spectral analysis of the frequency of characters of the original text, and Fig 3.2 shows the spectral analysis of the frequency of characters of the encrypted text.

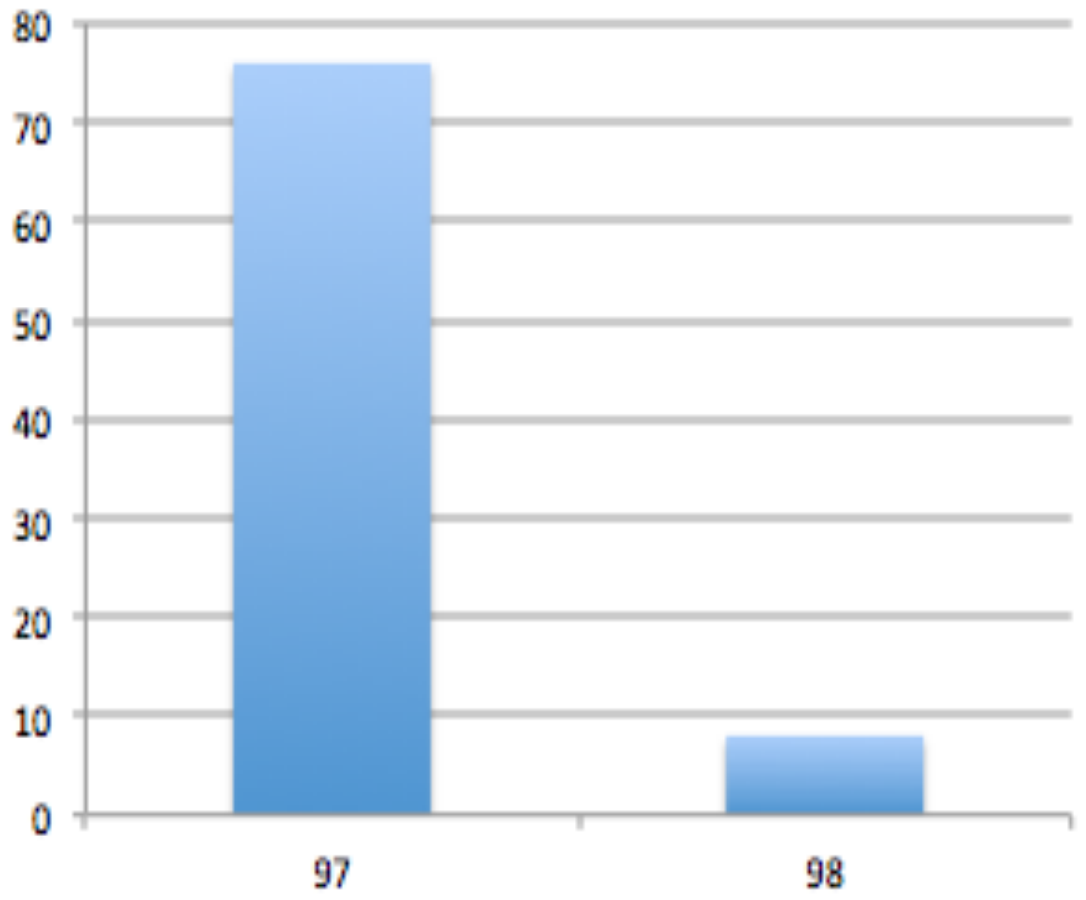

Fig 1.1: Spectral Analysis of frequency of characters of the palindrome of 76 ‘a’ and 8 ‘b’

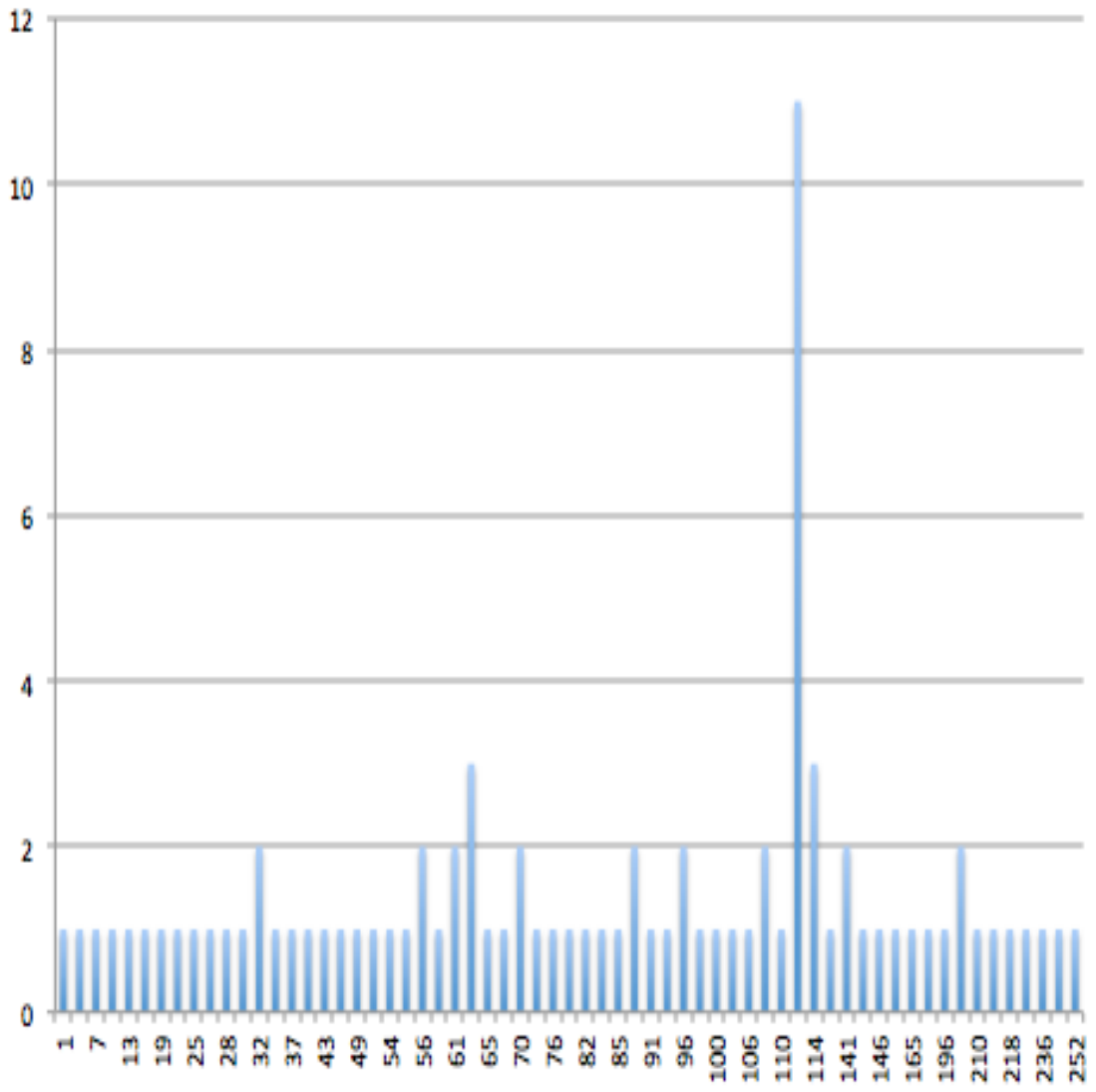

Fig 1.2: Spectral Analysis of the frequency of characters of the encrypted Palindrome of 76 ‘a’ and 8 ‘b’

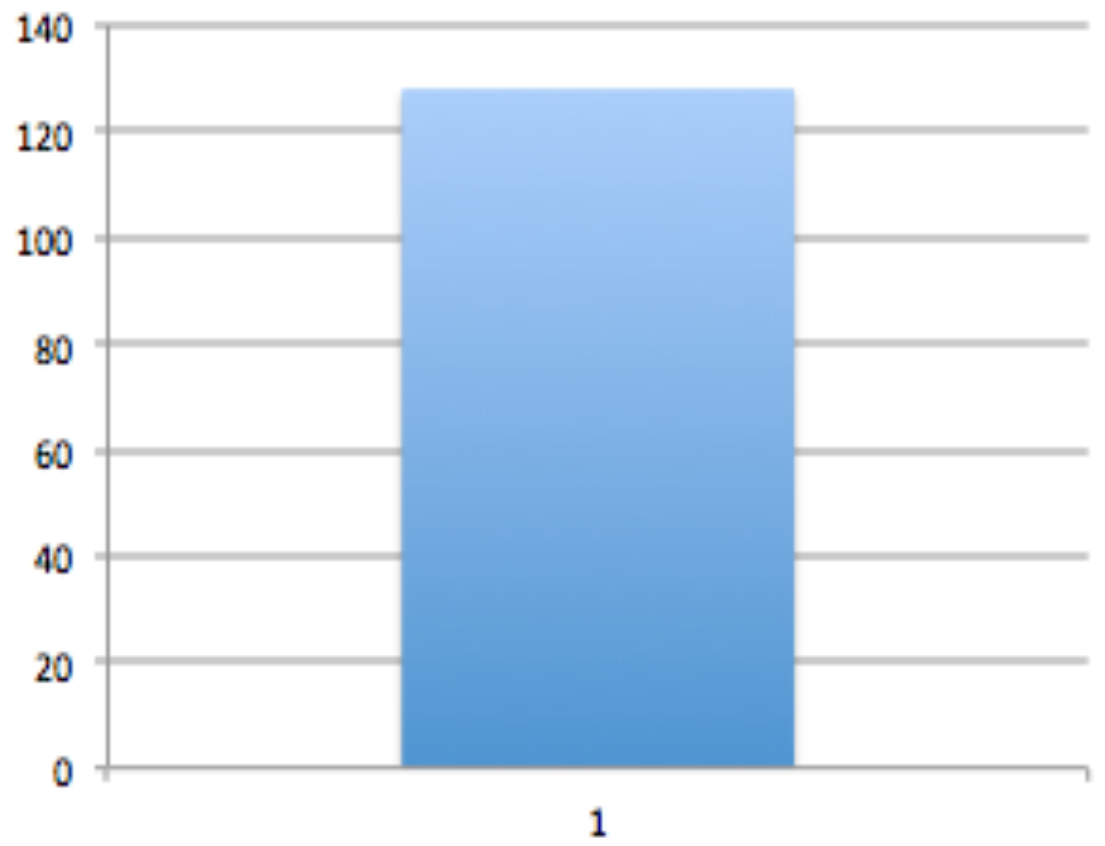

Fig 2.1: Spectral Analysis of the frequency of characters of 128 ASCII Value (1)

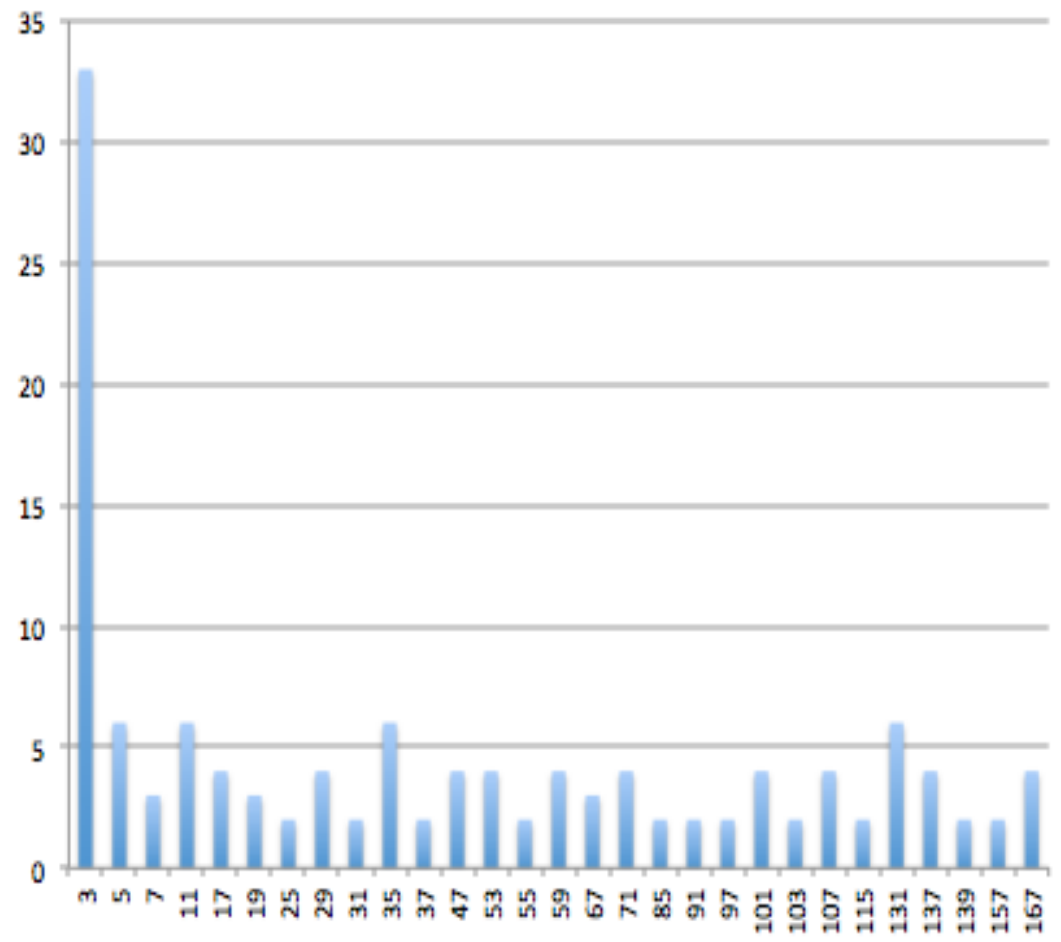

Fig 2.2: Spectral Analysis of the frequency of characters of the Encrypted text of 128 ASCII Value (1)

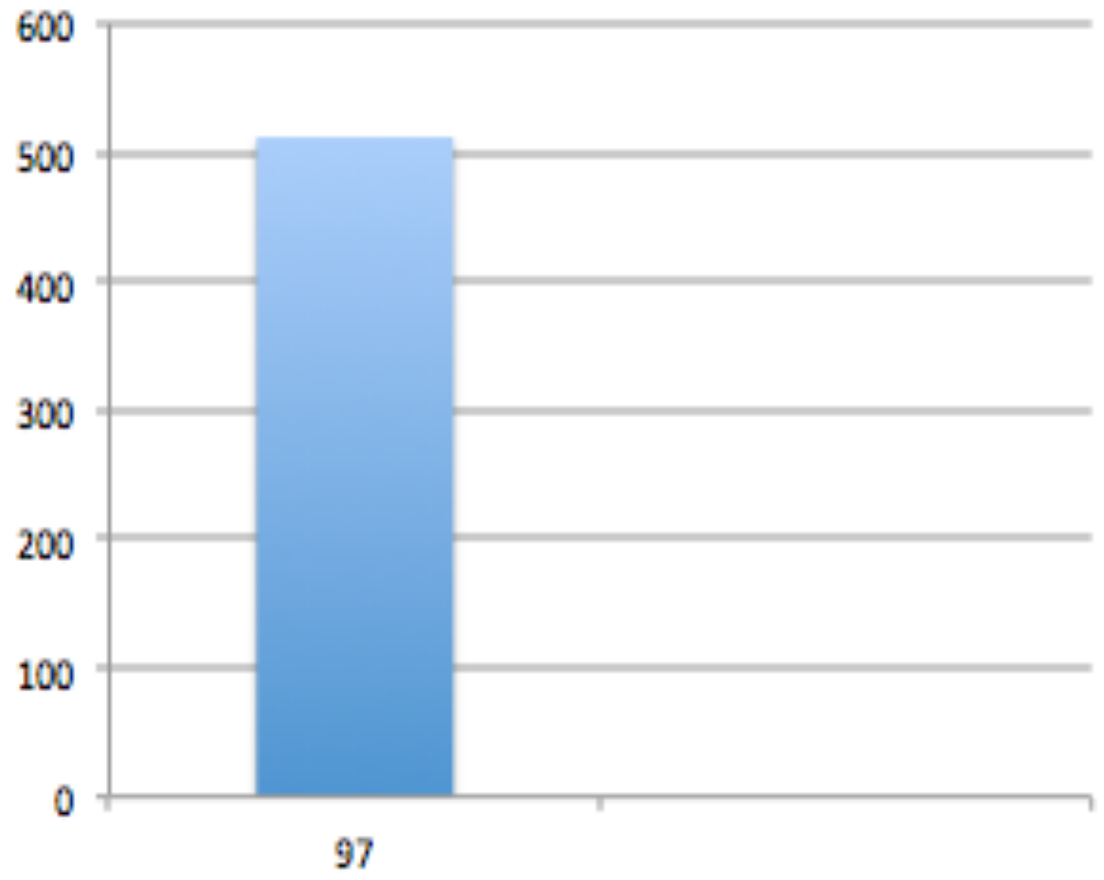

Fig 3.1: Spectral analysis of the frequency of characters of a text containing 512 times 'a'

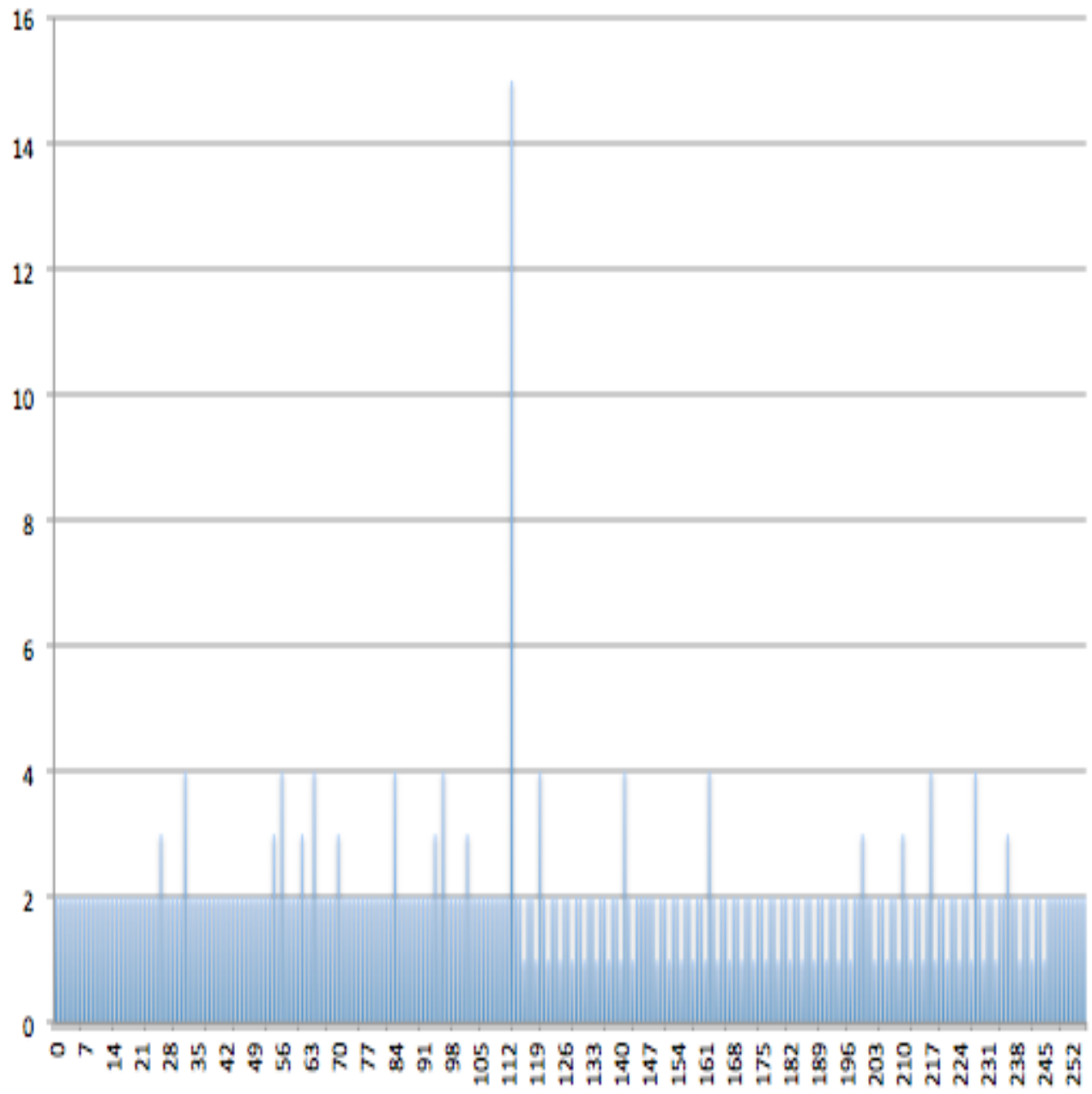

Fig 3.2: Spectral Analysis of the frequency of characters of the encrypted text containing 512 'a'

Thus, if we study the spectral analysis of each test case very carefully then we can see that there is no regular pattern or trace, which can be observed, to analyze the encryption technique or to retrieve the key. And the best part of SD-REE method is that the method is also very effective against palindromes, and not just repetitive bytes or terms. From the above given test cases, it also proves that SD-REE method is not limited by any ASCII value and works equally well for any character with ASCII value ranging from 0-255.

If we analyze the above SD-REE method, then we can also see that the use of polynomial function in the method have significantly increased the strength of encryption. Although this technique is based on the Caesar Cipher method but modifying Caesar Cipher by introducing random polynomial function and modular reduction have made a strong cryptographic method. This helps the encrypted text to be almost impossible to be detected by including repetitive bytes or characters and it also makes the method strong against Differential Attack (Differential Cryptanalysis).

## VI. Conclusion And Future Scope

Since, SD-REE cryptographic method uses modular-reduction [3,4], the system is cost effective, because the method executes computation of large integral values.

SD-REE method should be used with other encryption techniques present in the world, so that those encryption techniques can be made stronger and the possibility of breaking the encryption techniques or retrieving the key can be avoided by excluding repetitive terms from the encrypted text. In this way we can avoid the possibility of cryptanalysis and differential cryptanalysis (differential attack) occurring for an encryption technique. With the inclusion of bit-level encryption, SD-REE will be more effective and it will be almost impossible to break the cryptographic method or to retrieve the key from the encrypted text, and the author have already started working on that.

## Authors and Affiliations

*Somdip Dey (Author) is a student of B.Sc (Hons) Computer Science and affiliated to Department of Computer Science, St. Xavier's College [Autonomous], Kolkata, India.*

## Acknowledgment

Somdip Dey expresses his gratitude to all his fellow students and faculty members of the Computer Science Department of St. Xavier's College [Autonomous], Kolkata, India, for their support and enthusiasm. He also thanks Dr. Asoke Nath, professor and founder of Computer Science Department of St. Xavier's College (Autonomous), Kolkata, for his constant support and helping out with the preparation of this paper.